\documentclass[]{raa}
\usepackage{graphicx,times}
\usepackage{natbib}
\usepackage{amssymb,amsmath}
\usepackage{aas_macros}
\usepackage{color}
\usepackage[nodots]{numcompress}
\usepackage{verbatim}
\usepackage{url}
\bibpunct{(}{)}{;}{a}{}{,}

\usepackage[a4paper=true,dvipdfm=true,pagebackref=true]{hyperref}
\hypersetup{pdftitle = The title of my PDF, pdfauthor = My name, pdfsubject= The subject, pdfkeywords = keyword1 keyword2 keyword3}
\hypersetup{colorlinks = true, linkcolor = green, anchorcolor = red, citecolor = blue, filecolor = red, pagecolor = red, urlcolor = red}

\renewcommand{\vec}[1]{\boldsymbol{#1}}

\newcommand{\dif}{\mathrm{d}}

\def\eV{{\rm eV}}

\def\ns{{\rm ns}}

\renewcommand{\figurename}{Fig.}
\renewcommand{\tablename}{Tab.}

\begin{document}
\title{Numerical Simulation of Radio Signal from Extended Air Showers}

\volnopage{ {\bf 2014} Vol.\ {\bf X} No. {\bf XX}, 000--000}
\setcounter{page}{1}

\author{Wei Liu\inst{1, ~2}, Xuelei Chen\inst{1, ~3}}

\institute{National Astronomical Observatories, Chinese Academy of Science, Beijing 100012, China; {\it xuelei@cosmology.bao.ac.cn}\\
        \and
        University of Chinese Academy of Sciences, Beijing 100049, China; \\
        \and
Center of High Energy Physics, Peking University, Beijing 100871, China\\
\vs \no
{\small Received ; accepted }
}

\abstract{The burst of radio emission by the extensive air shower
provides a promising alternative for detecting ultra-high energy
cosmic rays. We have developed an independent
numerical program to simulate these radio signals. Our code
is based on a microscopic treatment,
with both the geosynchrotron radiation and charge excess
effect included. Here we make a first presentation of our
basic program and its results.
The time signal for different polarizations are computed, we find that the
pulses take on a bipolar pattern, the spectrum is suppressed towards
the lower frequencies. We investigate
how the shower at different heights in atmosphere contribute
to the total signal, and examine the signal strength and distribution
at sites of different elevations. We also study the signal
from showers of different inclination angles and azimuth directions.
In all these cases we find the charge excess effect important.
}

\authorrunning{L. Wei, X.-L. Chen}            
\titlerunning{Numerical Simulation of Radio Signal from Extended Air Showers}  
\maketitle

\section{Introduction}
\label{intro}
It is well-known that the high energy cosmic rays particles
could produce a large amount of secondary particles when they
enter the atmosphere through cascading reactions
with air molecules. These ensuing particles are called
\emph{Extensive Air Shower}(EAS). In 1965, the radio emission from
these shower particles were detected for the first time
\citep{1965Natur.205..327J}. This radio signal offers a way to detect the
very high energy cosmic ray. Later, more experiments were
carried out in order to further unravel the characters of this radio
signal. For a review of these early activities, see Ref. \cite{1971Allan}.
The radio detection technique has several advantages:  it can
operate round-the-clock with very little dead time, it is highly cost-effective,
hence very large effective collecting area can be achieved,
and it is sensitive to the atmospheric depth of the shower
maximum\citep{2010NIMPA.617..484H}. However, during the 1970s, as other
techniques matured and were considered more reliable at the time, the research
in this area dwindled.

In the last decade, with fast electronics and high-performance computers
appearing, there is a revival of interest in the radio detection of
cosmic rays air-showers. The ``LOPES''(LOFAR PrototypE Station)\citep{2005Natur.435..313F, 2013AIPC.1535...78S, 2013arXiv1303.7080A} in Germany and the
``CODALEMA''(COsmic ray Detection Array with Logarithmic ElectroMagnetic
Antennas)\citep{2009APh....31..192A} in France experimented with the
radio detection of very high energy cosmic ray particles, and
a new generation of radio detectors, called
the Auger Engineering Radio Array(AERA), is currently under construction
in the site of Pierre Auger Observatory in south
America\citep{2010NIMPA.617..484H, 2012NIMPA.662S.134S, 2011APh....34..717A}.
In the wake of LOPES success, cosmic ray detection appeared
on the agenda of the LOw Frequency ARray (LOFAR).
In Yakutsk, Russia, a radio arrays for similar purposes\citep{2013JPhCS.409a2070K}
have been built. A series of radio
experiments\citep{2011APh....34..717A, 2012NIMPA.662..S29M},
called ``TREND'', have been launched by a Sino-French team
in searching of ultra-high energy neutrinos, on the site of the 21cm
array(21CMA) radio telescope in Xinjiang, China.

The first prediction of radio emission from
the EAS was based on the idea that the extra electrons in the shower
could produce coherent C\'erenkov radiation at radio frequency
 \citep{1962JPSJS..17C.257A, 1965JETP...21..658A}. However,
\citet{1966RSPSA.289..206K} proposed that the geosynchrotron mechanism--the
synchrotron emission of electrons moving in the geomagnetic field-- as the
main source of the radio emission.
The radio pulses produced by the coherent geosynchrotron radiation
mechanism exhibits intense polarization effect, this has been confirmed
by recent experiments\citep{2009APh....31..192A, 2010APh....32..294A}.

In recent years, a number of different programs have been developed
to calculate the radio signal for a given cosmic ray shower.
In one approach, the radiation is calculated by assuming a ``macroscopic''
model of charge and current distribution in the shower
\citep{2008APh....29...94S, 2008APh....29..393W}.
The numerical computing program \emph{MGMR}\citep{2010APh....34..267D}
and \emph{EVA}\citep{2012APh....37....5W} have been
developed. In another, ``microscopic'' approach
the radio signal is computed by sampling the shower particles, and make a
coherent superposition of the synchrotron emission field of these
particles. The numerical program \emph{REAS} \footnote{http://www.timhuege.de/reas/}
was developed along this line\citep{2003A&A...412...19H, 2005A&A...430..779H,
2005APh....24..116H,2011APh....34..438L}.  Other models have also been
proposed, for example \emph{SELFAS}\citep{2012APh....35..733M, 2012arXiv1212.1348M}
and \emph{ZHAireS}\citep{2012APh....35..325A}.  The computations are
fairly complicated,
and there were very large differences in the predictions of these programs,
with the amplitude differ by as large as a factor of 20, and also
qualitatively in both the time domain (unipolar or bipolar pulse)
and frequency domain (flat or suppressed low frequency spectrum). Only
recently, after the charge excess effect have been included
in the computation with the ``endpoint formalism''\citep{PhysRevE.84.056602},
the numerical predictions of the various codes
begin to converge\citep{2012NIMPA.662S.179H}.

We have developed an independent numerical program to compute the radio
signals from the EAS. It is based on a microscopic model of the radio emission,
and both the geosynchrotron and charge excess effect have been included. While
the basic approach is to some extent similar to the
\emph{REAS} program, it is independently developed and many
details of the implementation is
different, hence it can furnish an independent check on the microscopic
approach. In this paper, we give an introduction to our formalism and
simulation program. We apply our program to study the characteristic
distribution of radio pulses and their dependencies on different
incident conditions, including the signal at different altitudes.
It will be the basis for a program of
further investigation on cosmic ray air shower radio emission.

This paper is organized as
follows: in section \ref{form} we
derive the electric field from shower particles, where both
geosynchrotron and radiation at the ends of particle's trajectory (charge
excess effect) are obtained and clearly distinguished. In section \ref{algo}
we describe our scheme of numerical simulation. In
section \ref{res}, the simulated results
are presented, where both time-domain signal and frequency spectra are
shown. We also study the contribution from the shower at different heights,
and give the result for observers at different elevations. We also consider
inclined showers and showers coming from different azimuth directions.
Finally we summarize our results in section \ref{con}.

\section{Radiation Formalism}
\label{form}
The canonical derivation of electric field of a moving charged particle
can be found in the standard textbooks\citep{Classical_Electrodynamics_Jackson,
walter1998classical, melrose2005electromagnetic}. The retarded potentials
produced by arbitrary-distributed sources are given by
\begin{eqnarray} \label{poten}
\phi(\vec{r}, t) &=& \frac{1}{4\pi \varepsilon_0} \int \dif t' \dif^3 \vec{r}'
\rho(\vec{r}', t') \frac{\delta(t-t'-|\vec{r} -\vec{r}'|/c)}{|\vec{r} -\vec{r}'|}\; ,\nonumber \\
\vec{A}(\vec{r}, t) &=& \frac{\mu_0}{4\pi} \int \dif t' \dif^3 \vec{r}' \vec{j}
(\vec{r}', t') \frac{\delta(t-t'-|\vec{r} -\vec{r}'|/c)}{|\vec{r} -\vec{r}'|}\; ,
\end{eqnarray}
where $\varepsilon_0$ and $\mu_0$ are respectively the permittivity and
permeability in free space, and $c$ is the speed of light in free space.
Here we neglect the deviation of refractive index from its
vacuum value (unity), and thus the \v{C}erenkov effect is neglected for
the present. $\delta(t-t'-|\vec{r} -\vec{r}'|/c)/|\vec{r} -\vec{r}'|$ is the
Green function of corresponding wave equation\citep{Classical_Electrodynamics_Jackson},
$\rho(\vec{r}', t')$ and $\vec{j}(\vec{r}', t')$ are
respectively the charge and current density of sources, and $|\vec{r} -\vec{r}'|$
gives the distance from source position $\vec{r}'$ to the observer position $\vec{r}$.

Charged particles are produced by pair creation or ionization at the shower front, and
then moves with the shower, contributing to the total radiation. After moving some
distance, they may lose their energy suddenly by major collisions, and left the shower.
The contribution to the radiation at both ends may be important and should be taken into
account.
The source term of a suddenly-created and destructed moving charge can be written as
\begin{eqnarray} \label{rho_den}
\rho(\vec{r}, t) &=&  e\delta^3(\vec{r} -\vec{x}(t)) \theta(t-t_s) \theta(t_e -t)\; , \nonumber \\
\vec{j}(\vec{r}, t) &=& e\vec{v}\delta^3(\vec{r} -\vec{x}(t)) \theta(t-t_s) \theta(t_e -t)\; ,
\end{eqnarray}
where $e$ is unit charge and $\vec{x}(t)$ is particle's trajectory in the geomagnetic
field. $\theta(t)$ is a Heaviside step function,
$t_s$ and $t_e$ respectively denote the starting and ending time
of the motion of a charged particle \citep{2012APh....35..733M}.
In order to integrate $\delta$-function in Eq.(\ref{poten}),
we introduce a new variable $u = t' +|\vec{r} -\vec{x}(t')|/c -t$, and beware of
$\dif u/\dif t' = 1-\vec{n}\cdot \vec{\beta}$,
the corresponding Lienard-Wiechert potentials can be obtained,
\begin{equation} \label{poten_res}
\phi = \left[ \frac{e}{4\pi \varepsilon_0 KR} \theta(t-t_{s}) \theta(t_{e}-t)  \right]_{ret},
\qquad
\vec{A} = \left[ \frac{\mu_0 e \vec{v}}{4\pi KR} \theta(t-t_{s}) \theta(t_{e}-t) \right]_{ret},
\end{equation}
where $K = 1-\vec{n}\cdot \vec{\beta}$,
and $R = |\vec{r} -\vec{x}(t')|$.
The quantities in the r.h.s have to be
evaluated at the retarded time $t'$, which is determined by the retarded
relationship $t = t' + R(t')/c$.
The electric field are evaluated in terms of the potentials by
$\vec{E} = -\nabla \phi -\frac{\partial \vec{A}}{\partial t}$,
then we have
\begin{equation} \label{E_res}
\begin{split}
\vec{E} =& \left\lbrace -\nabla\left[\frac{e}{4\pi \varepsilon_0 KR}
\right]_{ret} -\frac{\partial}{\partial t} \left[ \frac{\mu_0 e \vec{v}}{4\pi KR}
\right]_{ret} \right\rbrace \left[\theta(t-t_{s})\theta(t_{e}-t)\right]_{ret} \\
        \ \\
		&  + \left\lbrace -\left[ \frac{e}{4\pi \varepsilon_0 KR} \right]_{ret}
\nabla t' - \left[ \frac{\mu_0 e \vec{v}}{4\pi KR} \right]_{ret} \frac{\partial t'}{\partial t}  \right\rbrace \frac{\partial}{\partial t'}
\left[\theta(t-t_{s})\theta(t_{e}-t)\right]_{ret}\; .
\end{split}
\end{equation}
Here the first term is due to the continues motion of charge particles,
while the second term accounts for the sudden creation and
destruction.
Noting that\citep{citeulike:1637352},
\begin{equation} \label{eq:relationship}
\frac{\partial t}{\partial t'} = 1-\vec{n}\cdot \vec{\beta}\; , \qquad
\nabla t' = -\frac{\vec{n}}{c \cdot(1-\vec{n}\cdot \vec{\beta})}\; ,
\end{equation}
we have
\begin{eqnarray}
\nonumber \vec{E}(\vec{x}, t) &=& \frac{e}{4\pi \epsilon_0}\Bigg\lbrace
\left[ \frac{(\vec{n} - \vec{\beta})}{\gamma^2 K^3 R^2}\right]  _{ret}
\
	+ \left[ \frac{\vec{n}\times \{(\vec{n} -
\vec{\beta})\times \vec{\dot{\beta} } \}}{c K^3R} \right] _{ret}
\Bigg\rbrace \left[\theta(t-t_{s})\theta(t_{e}-t)\right]_{ret}
\ \\
	&& + \left[ \frac{e(\vec{n} -\vec{\beta})}{4\pi \varepsilon_0 K^2 R c}
\right]_{ret} \frac{\partial}{\partial t'} \left[\theta(t-t_{s})\theta(t_{e}-t)\right]_{ret}\; .
\label{eq:total_field}
\end{eqnarray}
In the braces, the first term is called the generalized Coulomb field and the
second term is the well-known radiation field, or acceleration field.
The third term indicates radiation from particle's creation and destruction. So in a neutral shower,
as both positive and negative charges (electrons and positrons) move toward ground,
the net contribution from third term is nearly zero. However, the
electrons from air molecules are knocked out by the cosmic ray and join the
shower, a real shower is negatively charged
\citep{1962JPSJS..17C.257A, 1965JETP...21..658A,2012APh....35..325A}.
This radiation from the excess electrons have a significant contribution to
the radiation, as we shall see below, and following others we shall call it the
charge excess effect.

\section{Algorithms}
\label{algo}

\subsection{Extensive Air Shower Properties} \label{distr}
The development of the air shower can be simulated with
Monte Carlo programs, such as CORSIKA\citep{1998cmcc.book.....H},
AIRES\citep{1999astro.ph.11331S} and COSMOS\citep{2013arXiv}.
As a first step, in this paper we apply parameterized functions to describe the
distribution of shower electrons and positrons, focusing on the
relation between the radio signal and shower properties, and
leave the detailed modelling of the air showers to future work.
Here we briefly recall the salient features of these
parameterized distribution functions used in the present simulation,
which were also used by \citet{2003A&A...412...19H, 2005A&A...430..779H}.

The ``Shower age'' $s$ is often used to mark the status of shower evolution,
and a good approximation of it is
\begin{equation}
s(X) = \frac{3X}{X +2X_{m}}\; ,
\end{equation}
$s$ varies between $1$ and $3$. $X$ is the atmospheric depth,
which is defined as an integral of air density along the shower path,
\begin{equation}
X(h) = \int_h^H \frac{\rho(h)}{\cos\theta} \dif h\; ,
\end{equation}
where $\rho$ is the atmospheric density, $H$ the initial height of shower
development and $\theta$ the zenith angle of the shower. $X_{m}$ signifies the
atmospheric depth where the shower reaches its maximum,
viz. $s = 1$, with
\begin{equation}
X_m= X_0 \ln (E_{p}/E_{c})\; ,
\end{equation}
 where $X_0 = 36.7$ g cm$^2$ is the radiation
length of the electron in the air, which is about $300$ m at sea level
and $E_{c} = 86$ MeV is the critical energy where the
ionization loss of the electron
equals to radiative loss. Below, as an illustration of the typical
case, we shall consider a cosmic-ray proton with primary
energy $E_p = 10^{17}$ eV. We model the atmosphere density$\rho(h)$
according to the US Standard Atmosphere 1977,
at layer $i$
\begin{equation}
\rho(h) = \frac{b_i}{c_i} \exp(-\frac{h}{c_i})\; ,
\end{equation}
where the parameters $b_i$ and $c_i$ for different layers
are listed in table \ref{tab:atmospherelayers}.

\begin{table}[htb]
\begin{center}
  \caption{
  \label{tab:atmospherelayers}
  Parameters for the parametrisation of the atmospheric layers(taken from \citep{2005A&A...430..779H}).}
  \begin{tabular}{ccccc}
     \hline
     Layer & Height [km] & $b_{i}$~[g~cm$^{-2}$] & $c_{i}$~[cm] \\
     \hline
     1 & 0 -- 4  & 1222.66 & 994186.38 \\
     2 & 4 -- 10 & 1144.91 & 878153.55 \\
     3 & 10 -- 40 & 1305.59 & 636143.04 \\
     4 & 40 -- 100 & 540.18 & 772170.16 \\
     \hline
  \end{tabular}
\end{center}
\end{table}

The profile of shower size, i.e. the total number of electromagnetic components
$N(s)$(including both electrons and positrons) at given shower age $s$ is
parameterized as \citep{1960ARNPS..10...63G}:
\begin{equation}
N(s) = \frac{0.31}{\sqrt{X_{m}/X_0}} \exp \left[\frac{X_{m}}{X_0} \frac{2 -3\ln s}{3/s -1}\right]\; .
\label{eq:Ns}
\end{equation}
And the number of particles injected per unit atmospheric depth $\vec{d}X$ is then
\begin{equation}
I(X) = \frac{\vec{d} N(s)}{\vec{d} X} + \frac{N(s)}{\lambda}\; ,
\label{eq:Ix}
\end{equation}
where $\lambda \approx 40$ g cm$^2$ is the mean free path of electron in the air.
The atmospheric depth of single particle follows a exponential distribution
$p(X) ~ \exp(-X/\lambda)$, and
$\frac{N(s)}{\lambda}$ is the annihilated particles per unit radiation length.

The lateral spread of shower particles  comes mainly from Coulomb scattering
of electrons off the air atoms. A favourite expression for the radial
distribution of electromagnetic components within the shower
is the Nishimura-Kamata-Greisen(NGK)
parameterization\citep{1958PThPS...6...93K, 1960ARNPS..10...63G}:
\begin{equation}
\varrho_{NKG}(r) = \frac{1}{r^2_{M}}\cdot \frac{\Gamma(4.5-s)}{2\pi \Gamma(s)\Gamma(4.5-2s)}\left(\frac{r}{r_M} \right)^{s-2} \left(1+\frac{r}{r_M} \right)^{s-4.5}\; ,
\end{equation}
where $r_{M}$ is the Moliere radius, which characterizes transverse spreading
of shower disk and a function of atmospheric depth $X$\citep{2003APh....18..351D}
\begin{equation}
r_M = \frac{9.6}{(X-a_i)}c_i\; .
\end{equation}
Showers developed at higher altitudes usually have wider lateral spread.

The thickness of the shower disk can be probed by measuring arrival time distribution.
A useful fitting formula is from \citet{1997APh.....6..301A}, which
contains both the curvature of the disk and the longitudinal distribution within it:
\begin{equation}
f(t) = At^B \exp{(-Ct)}\; ,
\end{equation}
where $t$ is the particle's arrival time at the detector relative to the shower front.
Parameter $A$ is a normalization, whereas $B$ and $C$ are functions of the
mean arrival time $\langle t\rangle$ and corresponding standard
deviation $\sigma_{t}$, both of
which are related to radial distance to the shower center,
\begin{align*}
& B = \left( \frac{\langle t\rangle}{\sigma_t}\right)^2 -1\; ,  \quad  C = \frac{\langle t\rangle}{\sigma^2_t}\; , \\
& \langle t\rangle(r),\  \sigma_t(r) = F + G (\frac{r}{r_0})^H\; ,
\end{align*}
where
\begin{align*}
F_t\  =&\  (8.039 \pm 0.068) \ns,  \quad   F_{\sigma} = (5.386 \pm 0.025) \ns\; , \\
G_t\  =&\  (5.508 \pm 0.095) \ns,  \quad   G_{\sigma} = (5.307 \pm 0.032) \ns\; , \\
H_t\  =&\  1.710 \pm 0.059,      \ \  \qquad  H_{\sigma} = 1.586 \pm 0.020\; ,
\end{align*}

The average energy of the electrons and positrons in the air shower is
about 30 MeV, where $\gamma \sim 60$\citep{1971Allan}. Following
\citet{2003A&A...412...19H}, we parameterize the energy distribution
of the cascading electrons as a broken power law, i.e.
\begin{equation}
p(\gamma) = \frac{\gamma}{74.2} \left( 1 - e^{-(\gamma/ 74.2)^{-3}}\right)\; ,
\label{eq:gamma_dist}
\end{equation}
where $\gamma$ denotes the Lorentz factor, which varies from $5$ to $1000$.
In this distribution, its maximum is at $\gamma = 60$.

A typical air shower is not neutral but have
more electrons, whose fraction usually
varies with atmospheric depth $X$ but have a mean of 23\%.
Here as a first approximation, we adopt a constant value of $25\%$.


\subsection{Strategy of Numerical Simulation}
\label{stra}

We use the Monte Carlo technique to simulate the radio emission.
Electrons and positrons are generated randomly according to the
shower distribution functions in a frame moving with the shower center, then
their positions in the ground reference frame are obtained by the coordinate
transformation (see Appendix A for details). The direction of
initial velocity is assumed to be along the radius of the spherical shower surface,
the subsequent motion of
the charged particles under the geomagnetic field is calculated according
to the Lorentz formula (See Appendix B), where we have neglected the energy
loss due to radiation or small angle scattering. We also use Monte Carlo to
determine the free path of each particle in order to determine where the
destruction take place. To take the radiation from the creation/destruction
of the charged particles into account, we need to estimate the number of particle
creations and destructions at each atmospheric depth, these are given
by the injection rate $I(X)$ as given in Eq.(\ref{eq:Ix}) and
destruction rate $|N(X)/\lambda|$.

\begin{figure}[tbp]
\begin{center}
\includegraphics[height=7.cm,angle=0]
{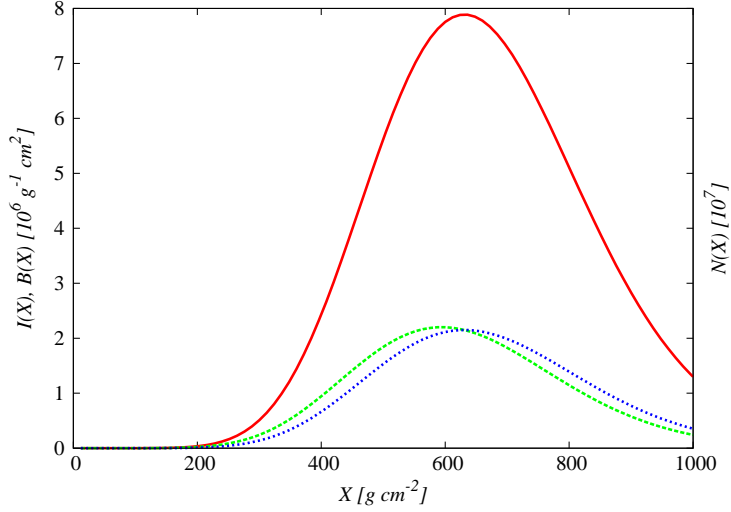}
\caption{
\label{fig:EAS_profile}
$N(X)$(red solid): the number of particles of an air shower as a
function of height, $I(X)$(green long dashed): number of injected particles per
unit atmospheric depth, $B(X)$(blue short dashed): number of
annihilated particles per unit atmospheric depth.
}
\end{center}
\end{figure}

In Fig.\ref{fig:EAS_profile}, we plot the number of particles $N(X)$,
injected particles $I(X)$ and the annihilated particles $B(X)$ at
different atmospheric depths in a
vertical shower(coming from the zenith).  The shower begin its development high in the
atmosphere, the number of particles increases as it moves downward, reaching a
maximum at $631 \mathrm{g~cm}^{-2}$ for a $10^{17}\eV$ cosmic ray proton,
i.e. about 4000 meters high, then the number of particles begin to
decrease. The injection rate $I(X)$ reaches maximum slightly
earlier than the total number
itself.

The time of emission and time of observation of the signal
are related by a nonlinear retardation relation.
Along the particle trajectory, a series of points are uniformly
sampled and their contribution to the electric field at the
corresponding observing time computed. We approximate the electric
field to be
$\bar{\vec{E}}(t_1) = \frac{1}{\bigtriangleup t}
\int^{t_1 +\bigtriangleup t}_{t_1} \vec{E}(t) \dif t$.
Here $\bigtriangleup t$ is the predefined time resolution, and for each
segment linear approximation is made. At both
endpoints of the trajectory, there are extra contributions from the creation
or destruction of the particle. In Eq.(\ref{eq:total_field}), the third term
reduces to
\begin{equation}
\left[ \frac{e(\vec{n}- \vec{\beta})}{4\pi \varepsilon_0 K^2Rc}
\left\lbrace \delta(t-t_{s})\theta(t_{e}-t) -\theta(t -t_s)\delta(t_e -t)
\right\rbrace \right]_{ret}\; .
\end{equation}
To get rid of the $\delta$-function, we can integrate 
for a very short interval, 
 $\int^{t_s+\epsilon}_{t_s-\epsilon} [...]_{ret} \dif t$, and the end point 
terms reduce to  
$\pm \left[ \frac{e~(\vec{n} -\vec{\beta})}{4\pi \varepsilon_0 KRc} \right]_{ret}\;$. 

For simplification, in our simulation we only generate electrons and positrons
which acquire velocity $\sim c$, but neglected the contribution from the 
positively charged ions which moves with much low speed. As 
$K=1-\mathbf{n}\cdot \mathbf{\beta}$, and the radiation term is proportional
to $K^{-1}$, this approximation is generally a good one. However, this 
omission could result in a longitudinal component of polarization when
calculating the end point radiation when the electron is ``created'' 
by ionization, or ``destructed'' by recombination, 
because it violates charge conservation at the 
creation and destruction point. This can be avoided by considering the 
contribution from the ion which are created or destructed at the same 
point. The velocity of such an ion is nearly zero, and the corresponding
end point radiation is   
$\left[ \frac{e~\vec{n}}{4\pi \varepsilon_0 Rc} \right]_{ret}$,
with the sign just opposite to the electron being created/destructed. So the
sum of the radiation along direction of observation is
\begin{eqnarray}
\nonumber \pm \left[ \frac{e~ (\vec{n} -\vec{\beta})}{4\pi \varepsilon_0 KRc} -\frac{e~\vec{n}}{4\pi \varepsilon_0 Rc} \right]_{ret} &=&
\pm \left[ \frac{e~ (\vec{n} -\vec{\beta} -(K = 1-\vec{n}\cdot\vec{\beta})\vec{n})}{4\pi \varepsilon_0 KRc} \right]_{ret}\; , \\
\nonumber &=& \pm \left[ \frac{e~ ((\vec{n}\cdot\vec{\beta})\vec{n} -\vec{\beta})}{4\pi \varepsilon_0 KRc} \right]_{ret}\; , \\
&=& \pm \left[ \frac{e~\vec{n} \times (\vec{n} \times \vec{\beta})}{4\pi \varepsilon_0 KRc}\right]_{ret}\; .
\label{eq:endpoint_trans}
\end{eqnarray} 
Then the radiation from charge excess effect only 
contain the part whose direction of electric field is perpendicular 
to the direction of observation. We shall use Eq.(\ref{eq:endpoint_trans}) to
calculate the end point radiation.

An actual shower of a $10^{17}$eV proton primary
have about $10^8$ shower particles, but in the Monte Carlo simulation only
a small fraction of these, usually a few million particles are sufficient.
We estimate the electric field as
\begin{equation}
\hat{\vec{E}} = \frac{N}{n} \sum_i^n \vec{E}_i\; ,
\end{equation}
where $N$ and $n$ are the expected total particle number and the sampled particle
number respectively. We use an adaptive control to
reach the required precision in sampling: in each iteration a batch of $10^5$
particles are added to the sample, and the electric field estimator at all the
required locations and time grid points are updated, and compared with the
value of last round. The number of location-time points where the relative
change exceeds the required precision ($10^{-3}$) is recorded.
Once such points are less than a predefined number, say 10 in 5000,
the results is considered to be stable and the simulation is terminated.
Our numerical program is implemented
using the C programming language with the aid of the Gnu
Scientific Library\footnote{http://www.gnu.org/software/gsl/}.

\section{Results}
\label{res}

\subsection{The contribution from different radiation mechanisms} \label{comp}

To understand how the different radiation mechanisms work, we calculate the
electric field signal from the pure geosynchrotron, the charge excess effect, and
their sum total. First we consider a $10^{17} \eV$
vertical shower and a $0.5$ Gs magnetic field
pointing due north horizontally. The shower is assumed to have an electron
excess of 25\%.

\begin{figure}[htb]
\begin{center}
\includegraphics[width=0.9\textwidth]
{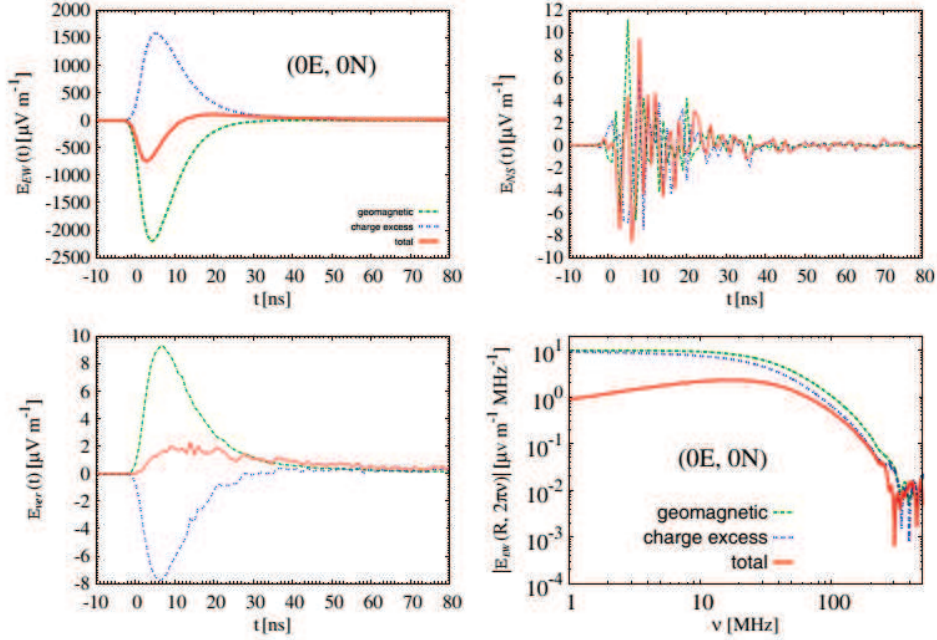}
\caption{\label{fig:signal_center} The three polarization signals (upper left: EW,
upper right: NS, lower left: vertical) and the frequency spectrum of the EW
polarization (lower right) as observed at the shower ground center,
with the pure geosynchrotron (green dash-dot curve), charge
excess effect (blue dash curve), and both (red solid curve).}
\end{center}
\end{figure}

\begin{figure}[htbp]
\begin{center}
\includegraphics[width=0.9\textwidth]
{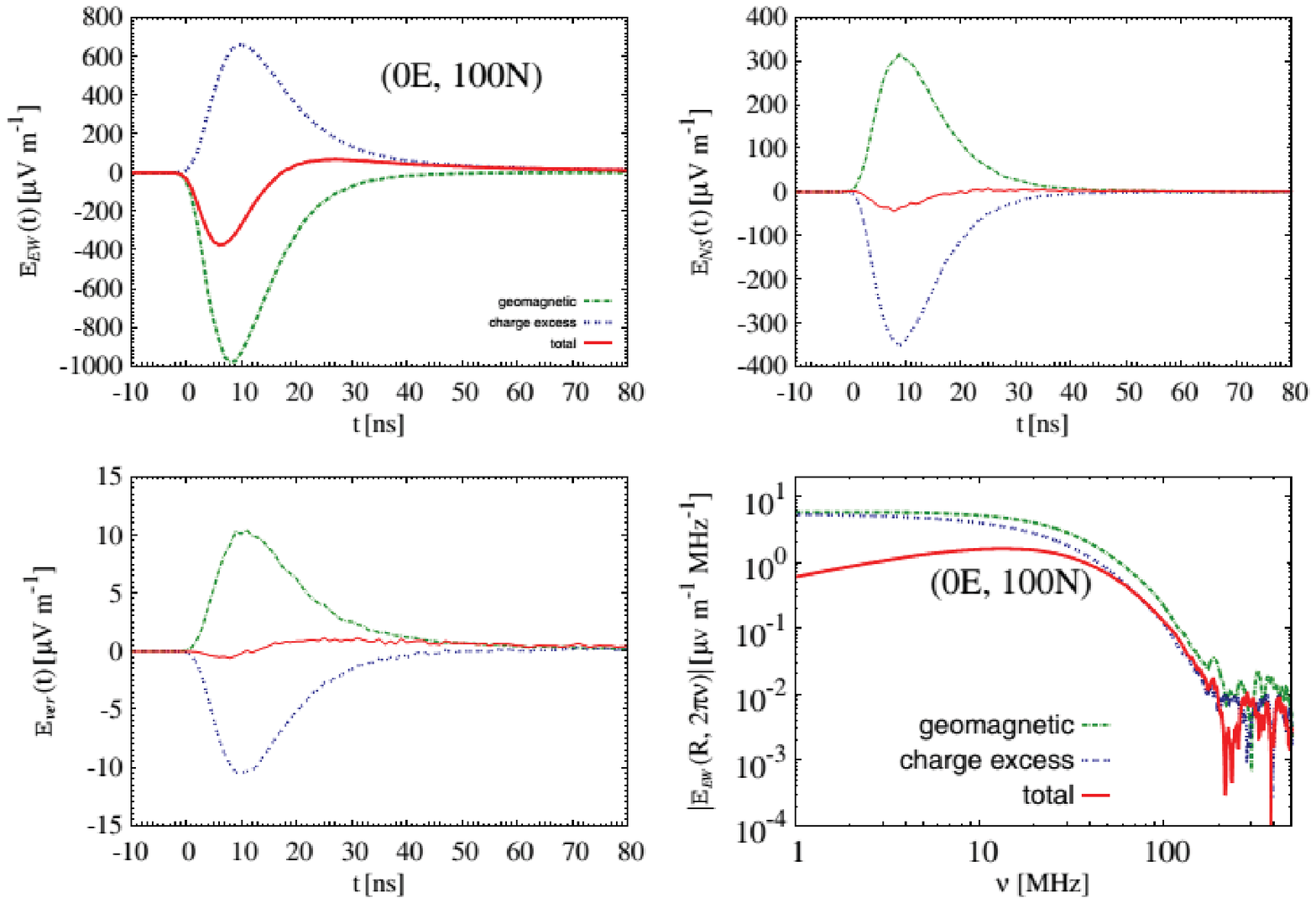}\\
\vspace{1cm}
\includegraphics[width=0.9\textwidth]
{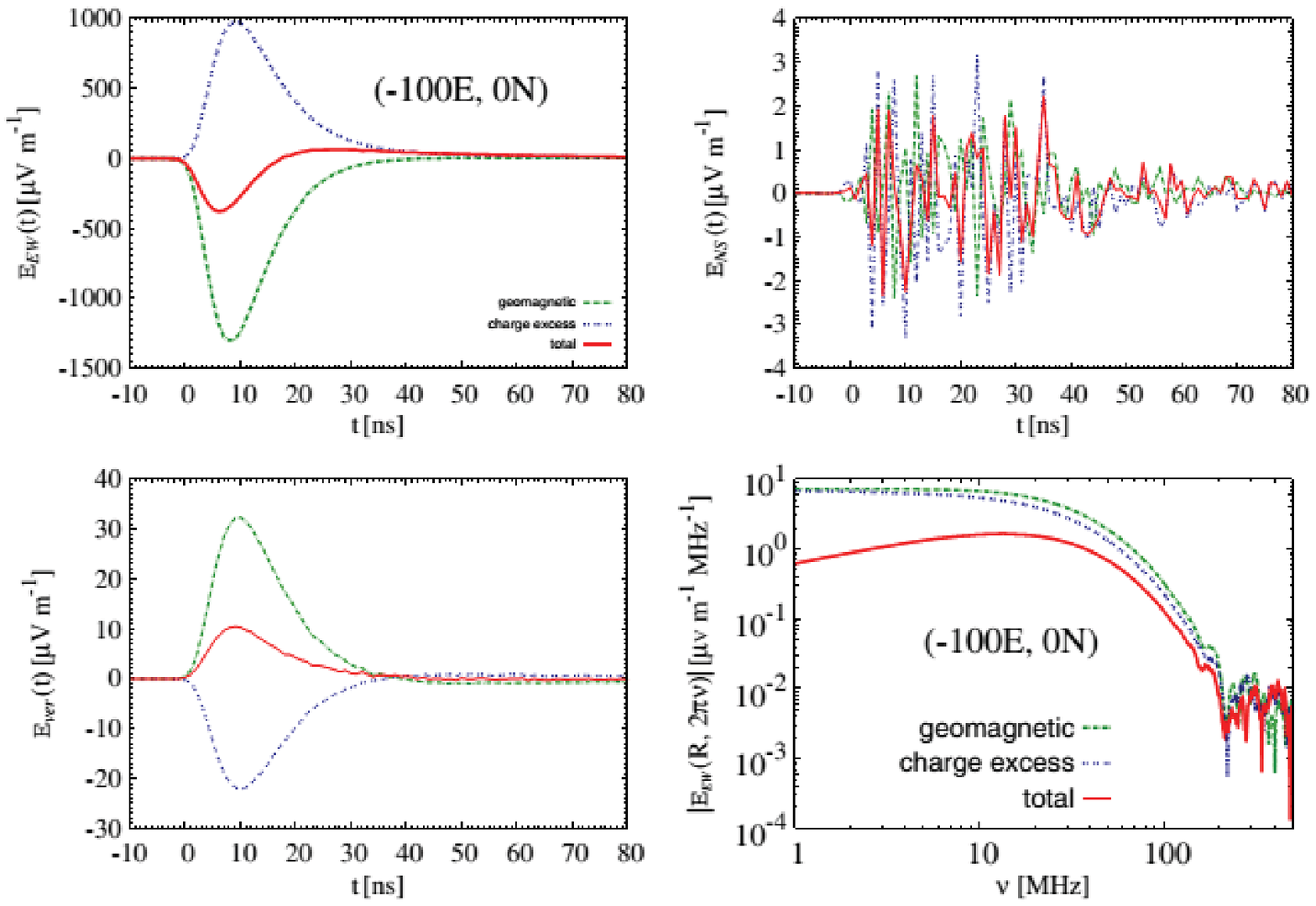}
\caption{\label{fig:signal_center_100} The same as Fig.~\ref{fig:signal_center},
except at off-center site. Top 4 panels: 100m north of the shower ground center, Bottom
4 panels:
 100m west of the shower ground center.}
\end{center}
\end{figure}

The electric field signal at the ground impact center of the shower axis
is shown in  Fig. \ref{fig:signal_center}, with polarizations in the East-West (EW)
direction, North-South (NS) direction, and vertical direction, as well as the
frequency spectrum for the signal.
Under the Lorentz force from the geomagnetic field, the charged particles in the
vertical shower are deflected toward east and west, as a result, we expect a
linear polarization in the geosynchrotron radiation in the East-West (EW)
direction, while the North-South (NS) polarization is expected to be small,  and the
vertical polarization is expected to vanish as it is along the line of sight.
These expectations are confirmed in Fig.\ref{fig:signal_center} where a strong
pulse in the EW polarization due to the pure geosynchrotron mechanism is
shown as the green dash-dotted line in the negative (west), which peaks at $8$ ns, with
a strength of almost $2000\mu$Vm. The NS polarization oscillates with small amplitude,
while the vertical polarization vanishes.

However, when the charge excess effect is included,
we see it makes prominent and opposite contribution to the total electric field,
shown as the blue dotted curve. As a result, it cancels
a large part of the field generated by the geosynchrotron mechanism,
especially for the primary EW polarization. The net effect, shown as the red solid
curve, is a much reduced pulse, of only about $400\mu$Vm at its peak, and even a
bipolar character where the signal at the later time is reversed in sign
from the earlier one, which is different from the unipolar pattern under
the pure geosynchrotron radiation. Whether the pulse is unipolar or bipolar
have been debated and it was only recently resolved that the difference
is due to the inclusion of the charge excess effect \citep{2012NIMPA.662S.179H}.
There is also a slight
vertical component at the level of $\sim 2 \times 10^{-2}$ of
the total signal, probably due to the finite size of the shower disk, and also
due to the asymmetry in charge.

Next we consider the signal at off-center locations. In Fig.\ref{fig:signal_center_100},
we plot the signals at a site 100m due north of the ground center (top 4 panels),
and a site 100m west of the ground center (bottom 4 panels).
Again, many of the basic features are similar to the case in the ground center, with the
EW polarization still the dominant one, though the amplitude is smaller than at the ground
center. In the off-center case, the NS polarization may be present, but
interestingly, in the 100m north case, both the the pure synchrotron and the
charge excess effect alone could produce relatively large peak, but they nearly
cancel each other and the net effect is a relatively small peak.

\begin{figure}[tbp]
\begin{center}
\includegraphics[height=0.6\textwidth]
{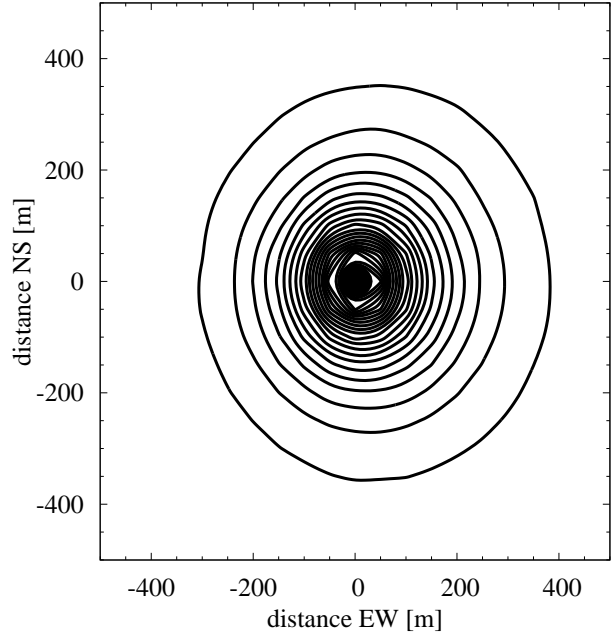}
\caption{
\label{fig:asycont}
The contour of electric field maximum of EW polarization
from a vertical shower. The contour levels are $25 \mu V m^{-1}$ apart.
}
\end{center}
\end{figure}

The whole pattern of the shower signal is shown in Fig.\ref{fig:asycont}.  The
signal is highly beamed, and we can see there is a slight asymmetry
in the EW direction. The shower is nearly vertical, but the Lorentz force
deflects motion of particles, and there is a net charge excess in the shower, in the end
it produced the pattern as shown in Fig\ref{fig:asycont}.

\subsection{Frequency Spectra Fitting}
\label{fre_dep}

In Fig.\ref{fig:signal_center} and Fig.\ref{fig:signal_center_100}, we have also
plotted the frequency spectrum of the radiation for different mechanisms (bottom right
panels in each of the four-plot combination).
The major component of the radio emission lies in the frequency range of
about tens of MHz, in agreement with observations.

At the high frequency end, we see from these figures that for
both the geosynchrotron and the charge excess effect,
the spectra fall off exponentially above $\sim 100$MHz, though the charge excess
radiation decays slightly earlier than the geosynchrotron radiation.
As a result, the total spectra also falls off. This spectral drop off
is due to the loss of coherence, because at such
high frequencies the wave length is far less than the thickness of the shower,
and the contribution to the field strength from different part of the shower no longer
simply adds up. As a result, the radiation is insignificant at such high frequencies.

At the low frequency end, we can see from these figures that
for both the geosynchrotron and charge excess effect  the spectra is nearly flat.
However, as the two are in opposite direction, they cancel each other,
and the total signal have a slowly decreasing
spectrum at the lower frequencies. This cut off at the lower frequency due to the
cancellation effect has been noted in the recent
literature\citep{2008APh....29..393W, 2011APh....34..438L,
2012APh....35..733M, 2012APh....35..325A}.

Experimentally,  analyses of a few strong events by the
CODALEMA\citep{2006APh....26..341A}
and LOPES\citep{2008A&A...488..807N, 2009NuPhS.196..297H} experiment show
that in the range of $30-70$ MHz, the frequency spectrum can be
well fitted with a single power-law $\epsilon_{\nu} = K\cdot \nu^{-\alpha}$
with spectral index $a = -1 \pm 0.2$, or alternatively by an
exponential function $\epsilon_{\nu} = K\cdot \exp(\nu/\mathrm{MHz}/\beta)$ ,
where $\beta$ spans from $-0.021$ to $-0.013$. This is slightly steeper
than the slope predicted by the pure geosynchrotron. In these
experiments, they found no significant dependence of the spectral slope on the
distance to the shower axis, the zenith angle or the
azimuth angle.

\begin{figure}[htbp]
\begin{center}
\includegraphics[width=0.6\textwidth]{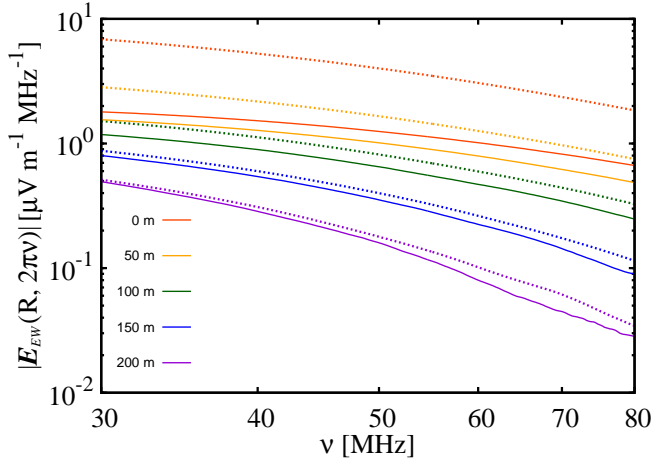}
\caption{
\label{fig:spec1}
The frequency spectra at different radial distances to the east of shower
ground center. The distances are respectively $0$ meter, $50$ meters, $100$
meters, $150$ meters, $200$ meters. Short dash lines: only geosynchrotron radiation;
Solid lines: with both geosyhchrotron and charge excess effect.}

\end{center}
\end{figure}

\begin{table}[htbb]
\caption{
\label{tab:spec_fit}
The fitted parameters of the frequency spectrum at different radial distances along
the east and west direction. We fit with a single power
law $\epsilon_{\nu} = K\cdot \nu^{-\alpha}$ between $40-70$ MHz.
}
\centering
\begin{tabular}{c c c c c c}
\hline\hline 
Distance & Orientation & $E_{0}$ (pure) & $\alpha$ (pure) & $E_{0}$ (both) & $\alpha$ (both)
\\ [.5ex]
\hline 

0 m & center & $1114.72$ & $1.442$& $96.37$ & $1.114$ \\[2ex]

& east  & $1114.72$ & $1.442$ & $161.56$ & $1.301$ \\[-1ex]
\raisebox{1.5ex}{50 m} & west \raisebox{1.5ex}
& $1403.90$ & $1.565$ & $132.16$ & $1.268$ \\[2ex]

& east  & $573.62$ & $1.680$ & $520.54$ & $1.714$ \\[-1ex]
\raisebox{1.5ex}{100 m} & west \raisebox{1.5ex}
& $2836.44$ & $1.893$ & $410.51$ & $1.681$ \\[2ex]

& east  & $2425.63$ & $2.233$& $3846.32$ & $2.383$ \\[-1ex]
\raisebox{1.5ex}{150 m} & west \raisebox{1.5ex}
& $10454.02$ & $2.405$& $3152.25$ & $2.375$ \\[2ex]

& east & $17889.64$ & $2.952$ & $110199.9$ & $3.453$ \\[-1ex]
\raisebox{1.5ex}{200 m} & west \raisebox{1.5ex}
& $50671.17$ & $3.012$ & $93022.37$ & $3.464$ \\[1ex]

\hline 
\end{tabular}
\end{table}

In Fig. \ref{fig:spec1} we plot the simulated spectra
at different distances from the ground center for a vertical shower.
Short dash lines are the spectra for the pure geosynchrotron, while
solid lines are those including charge excess effect.
The distances to shower impact center are respectively $0$ meter,
$50$ meters, $100$ meters, $150$ meters, $200$ meters.
We also fit these spectra with a single power law
$\epsilon_{\nu} = E_0 \cdot \nu^{-\alpha}$ in the range
of $40-70$ MHz, the fitting values
of $E_0$ and $\alpha$ are reproduced in table \ref{tab:spec_fit}, for
distances measured to both the east and the west of the ground center, as there is
a slight asymmetry as we noted before.
It is apparent that within $150$ meters, the
single power law is a good fit to the spectrum. With the radial
distances increasing, the slope becomes steeper. Near the center (within
50 m) part, the slope of the spectrum including both geosyhchrotron and
charge excess effect is $-1 \pm 0.2$, consistent with the experiment
result. On the other hand, for the pure
geosynchrotron radiation the slope $\alpha$ conflicts with the experimental
results, and the difference is larger than the margin of error. This shows that
the inclusion of the charge excess effect is very important.

However, far away from the center, both the pure geosynchrotron model and the
model with charge excess effect predict steepening of the spectra, while
observations so far have not found such change. Part of
this may be due to experimental error, because far from the shower
ground center, the signal strength falls off exponentially, and the
resulting measurement error is large. Another possibility is,
the \v{C}erenkov radiation may have visible effect at intermediate
distances \citep{2012NIMPA.662S.175D}, and may
modify the corresponding frequency spectrum.

\subsection{Contribution from different Elevations }
\label{evol}

We now study how the shower at different heights contribute to the
total signal at the ground level. If the shower is point-like,
there would be a one-to-one relation between the emission time
and arrival time for the radio pulse,
with the radiation emitted earlier arrive earlier, and
the envelope of the signal would clearly reflect that
of the shower at different heights\citep{2012APh....35..325A}.
However, the real case is more complicated due to the spatial
extent of the shower disk, for at any given time the signal received on
a location in the ground (altitude 0) is a superposition of emissions
from different parts of the shower at different times.
The problem of contribution from different heights in the case of
pure geosynchrotron was investigated by \citet{2007APh....27..392H},
here we consider the case with charge excess effect.

\begin{figure}[tbp]
\begin{center}
\includegraphics[height=6.cm,angle=0]{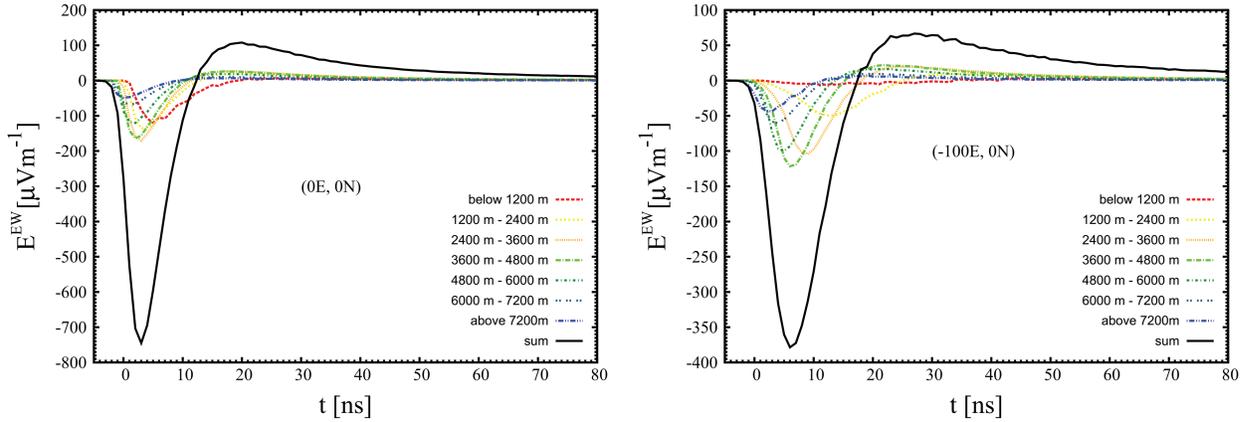}
\caption{
\label{fig:E_evolu_new}
Left: Contribution from layers at different heights at
the shower ground center.
Right: Contribution from layers at different heights at $100$m west from the shower ground center.
}
\end{center}
\end{figure}

We compute the radio signal contribution
from different height layers, the results are shown in Fig. \ref{fig:E_evolu_new}.
 For illustration, we have chosen two observing sites, one
at the shower ground center and one at $100$m-west off center.

Starting from the ground, we divide the whole atmosphere into
$7$ segments, each segment have an intervals of $1200$ m, except for the highest one,
where we combine all contributions from above  $7200$ m. We plot the contribution of each
layer as well as the total signal. A general impression is, each layer can contribute both
to the first, strong (negative) peak as well as the second, weak (positive) peak, though
the higher layers contribute more to the first while the lower ones contribute more to the
second. The $3600-4800$ m and $2400-3600$m layers contribute
the largest signal, these two are also the layers which contain the maximum number of
particles. The contributions from the higher elevations are smaller but still significant.
The contribution of the lower layers ($0-1200$m and $1200-2400$m) are also sizeable, they
are closer to the observer, but the number of particles have decreased. In particular,
the contribution of $0-1200$m is sizeable in the ground center, but away from ground center
it is much less, due to the relativistic beaming.

\begin{figure}[htbp]
\begin{center}
\includegraphics[height=6.cm,angle=0]{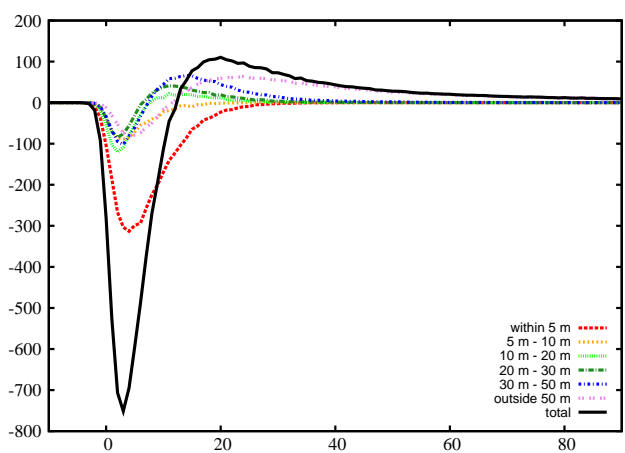}
\caption{
\label{fig:lateral_con}
Contributions at different lateral distances to shower axis.
The intervals are within $5$ meters, $5 - 10$ meters, $10 - 20$ meters,
$20 - 30$ meters, $30 - 50$ meters, and beyond $50$ meters.
}
\end{center}
\end{figure}

We further compute the contribution from different lateral distances,
see Fig. \ref{fig:lateral_con}. The observing site is chosen to be the shower
center. We make concentric rings around the shower axis, first ring within $5$m,
then $5 - 10$m,  $10 - 20$ m, $20 - 30$ m, $30 - 50$ m, and beyond $50$ meters, and
estimate the contribution from each. The main contribution comes from
the distance within $50$ meters, especially within
$5$ m. This is because most of the shower particles are located in the
inner rings near the center of the shower, as
the horizontal motion caused by geomagnetic
field is small compared with shower velocity.

\subsection{Signals for observing at different elevations}

The particle based cosmic ray detectors are often placed at sites of
high altitudes, because
the shower maximum is high in the atmosphere, and radio detectors
may also be located on the same sites, so it is important to consider the
altitude effect on the radio signal.
As we get closer to the shower maximum, we may receive stronger
emission from this stage. On the other hand, at the higher
elevation, the signals from the later stage of the shower is missed.

\begin{figure}[htbp]
\begin{center}
\includegraphics[height=6.cm,angle=0]{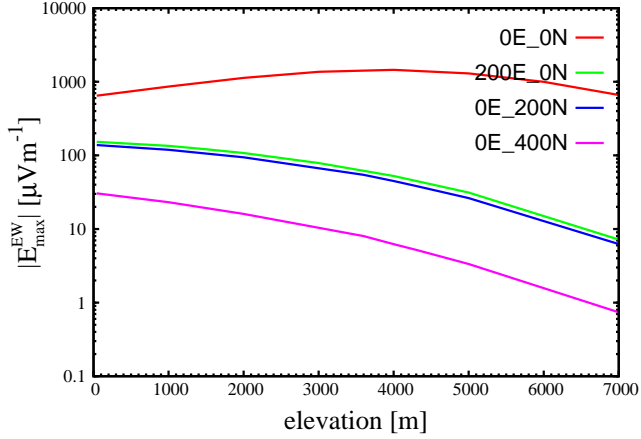}
\caption{
\label{fig:E_eleva}
Elevation dependence of the radio signal. Signals at four locations (ground center,
200m due East to the center, 200m due North to the center, 400m due North to the center)
are plotted at as a function of ground altitude.
}
\end{center}
\end{figure}

\begin{figure}[tbp]
\begin{center}
\includegraphics[height=5.cm,angle=0]{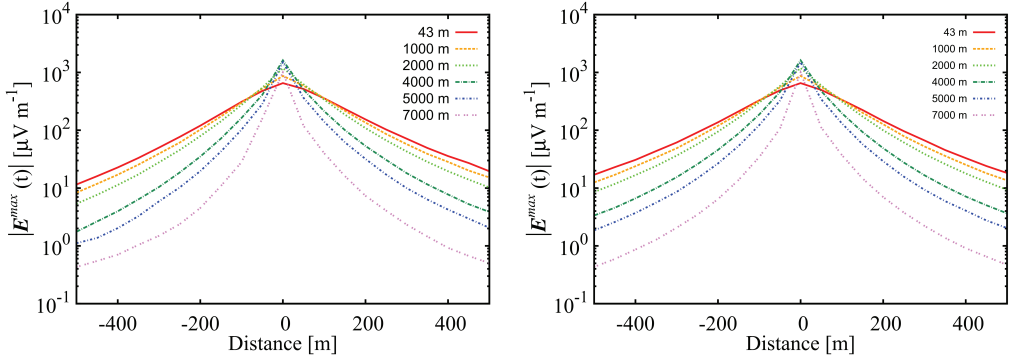}
\caption{
\label{fig:E_radial}The peak field strength of the pulse as a function of
off-center distance in the EW (top) and NS (bottom) directions.
}
\end{center}
\end{figure}

\begin{figure}[htbp]
\centering{
\includegraphics[height=6.cm]{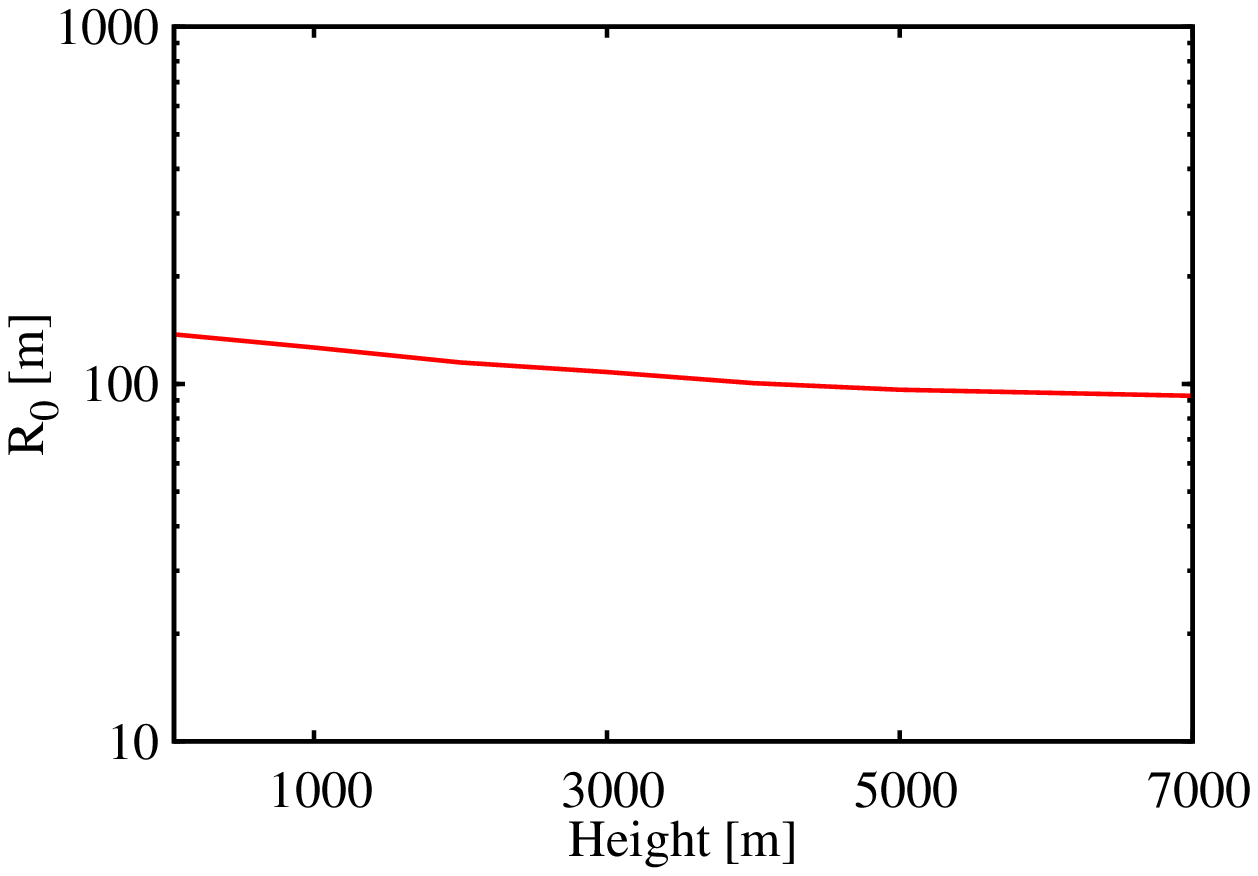}
\caption{\label{fig:R0_eleva} Elevation dependence of the scale $R_0$. }
}
\end{figure}

Fig. \ref{fig:E_eleva} shows how the peak field strength of the
radio pulse varies with observation site elevation for
a vertical shower at several offset differences from the ground center.
In all cases the variation is apparent but not very rapid.
In the case of shower ground center, the
signal strength raises gradually at the beginning and reaches its
maximum value at around $4000$ meters high, where the shower
develops to its maximum for a $10^{17}$ eV cosmic ray. At still higher
altitude, the signal begin to attenuate. In the off center cases,
the peak strength drops off with increasing altitude, and for the
three off-center distances we computed, the variations have similar
dependencies on height. This result show that if the radio detector array is
primarily designed primarily to detect the signal in the center area, then
there is a little advantage to choose a site of high altitude, though it is
far less significant as in the case of particle detectors. On the other hand,
if the radio array is sensitive enough to detect showers outside the center
area, then perhaps there is not much advantage to place the detector on high
altitudes.

In Fig. \ref{fig:E_radial} we plot the lateral
distribution of the radio signal at different elevations
for a vertical air shower. There is a slight asymmetry
of lateral distribution along the EW direction about shower axis,
where the signal in the east is stronger. This is caused by the
excess of electrons in the shower, but this does not affect the
NS distribution. Close to the shower axis, the peak electric field
strength raises with the increase of elevation until about $4000$m,
where the shower reaches the maximum for a $10^{17}$ eV cosmic
ray primary. Away from the shower axis, the strengths always
decreases  with the height. The turning point between the
{\it center} and {\it off-center} is at around $50$ meters.
The radial dependence of electric field signal is
usually parametrized by an exponential function
\begin{equation} \label{fit}
\varepsilon(r) = \varepsilon_{100} \exp(-(r-100\mathrm{m})/R_0)\; ,
\end{equation}
where $\varepsilon_{100}$ is the amplitude at $100$ m and scale
parameter $R_0$ is usually about $100$ to $250$ meters,
except for some events which has a very
large $R_0$\citep{2009NuPhS.196..297H, 2010APh....32..294A}.
We use Eq.(\ref{fit}) to fit the lateral distribution in the range
of $200-500$m at different elevations, and show the variations
of $R_0$ with heights in fig. \ref{fig:R0_eleva}. We can see that
$R_0$ do not change significantly with height.

\subsection{Inclined showers: zenith angle and azimuth angle dependence}
\label{zenith_dep}

\begin{figure}[htbp]
\begin{center}
\includegraphics[height=6.cm,angle=0]{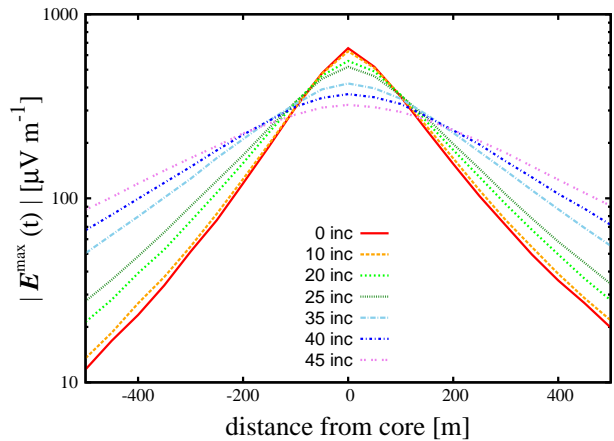}
\caption{
\label{fig:E_radial_zenith}
The radial distribution of the magnitude of radio signal under different zenith angles. }
\end{center}
\end{figure}

So far we have been considering only vertical showers, but inclined showers are of course
more common. The inclined showers share some general characters with the
vertical ones, now we will investigate how the shower behaviour changes as the
inclination angle varies.
The zenith dependence of radial distribution is shown in
Fig. \ref{fig:E_radial_zenith}. At the shower center, the field
strength decreases with zenith angle, the vertical shower have the
largest peak strength. However, the inclined showers have broader distribution,
so some distance away from the center they may actually have greater
field strength.

\begin{figure}[htbp]
\begin{center}
\includegraphics[height=15.cm,angle=0]{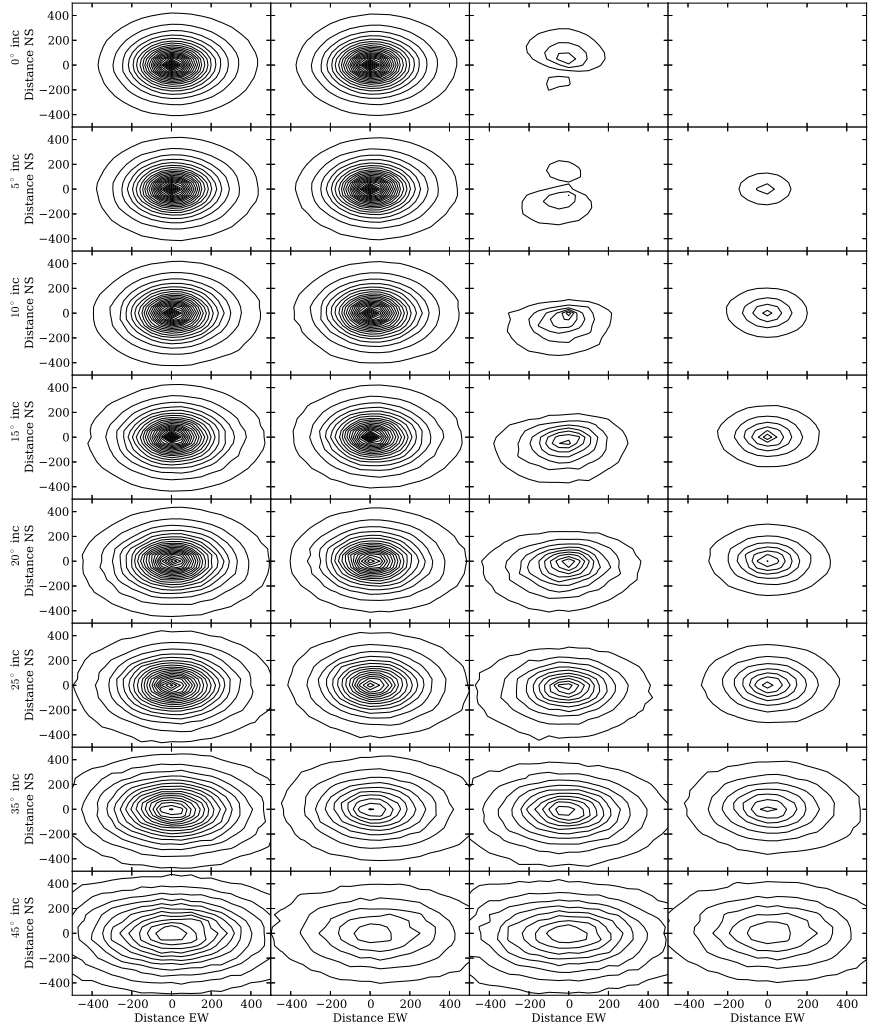}
\\
\caption{
\label{fig:E_contour_zenith}
The contours of unfiltered radio emission under different zenith angles.
Columns from left to right: total field strength, the
EW polarization, the NS polarization, and the vertical polarization.
Lines from top to bottom are with zenith angle of
$0^{\circ}$, $5^{\circ}$, $10^{\circ}$, $15^{\circ}$, $20^{\circ}$,
$25^{\circ}$, $35^{\circ}$, $45^{\circ}$ respectively.
Contour levels are $25 \mu \mathrm{V m}^{-1}$ apart.
}
\end{center}
\end{figure}

In Fig. \ref{fig:E_contour_zenith}, we show the contours of
radio emission field strength with different zenith angles, where
the shower axis is assumed to be inclined from the east
direction with different angles. Such spatial distribution could be
detected with a phalanx of radio signal receivers, and we show the
distribution for the total strength as well as the polarized
electric field along the EW, NS and vertical directions.

\begin{figure}[p]
\begin{center}
\includegraphics[height=15.cm,angle=0]{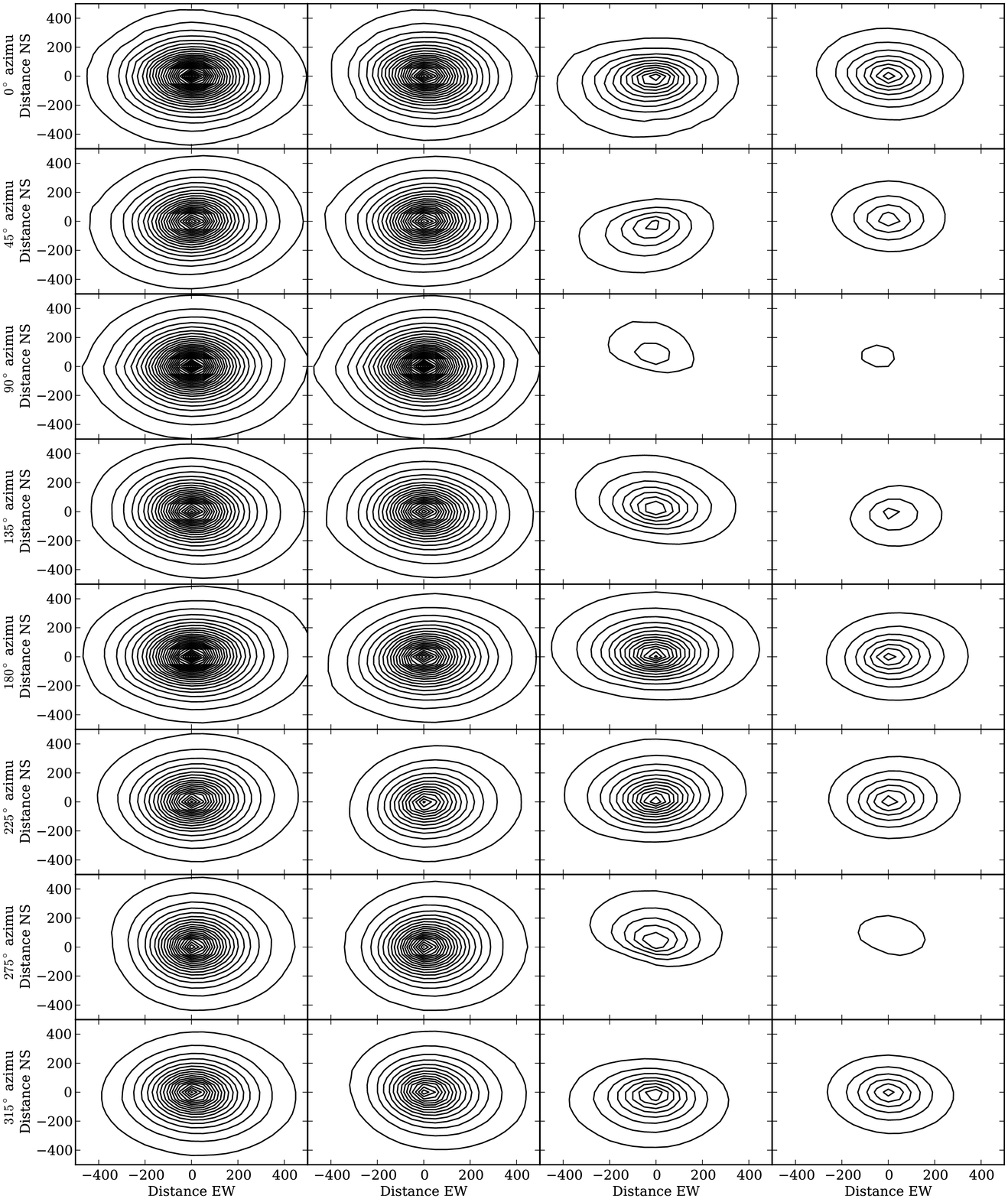}
\\
\caption{
\label{fig:E_contour_azimu}
The contours of unfiltered radio emission for a $20^{\circ}$-inclined
shower coming from different incident azimuthal directions.
Columns from left to right are respectively the total field strength, the
EW polarization, the NS polarization, and the vertical polarization.
Lines from top to bottom are with azimuth angle of
$0^{\circ}$, $45^{\circ}$, $90^{\circ}$, $135^{\circ}$,
$180^{\circ}$, $225^{\circ}$, $270^{\circ}$, $315^{\circ}$ respectively.
Contour levels are $25 \mu \mathrm{V m}^{-1}$ apart.}
\end{center}
\end{figure}

For a  vertical shower (zenith angle $0^\circ$),
the total field strength and the dominant
EW polarization components have a distribution of
concentric ellipses. The NS and vertical polarizations, on the other hand,
exhibit asymmetric bivalve structures in this case.
This asymmetry is due to the effect of magnetic field, which breaks the
otherwise totally symmetric arrangement in the vertical shower.

With increasing zenith angles, the total field strength and
the dominant EW polarization decreases slightly and their contour ellipses
become more prolate along the EW direction, and the axis of the contour ellipses
become longer. At the same time, the bivalve structures in the
NS and vertical components change to concentric ellipses and now these
components have greater magnitudes than the vertical case.
their magnitudes gradually grows. These changes are consistent with our
expectation for a slanted grant cross section with respect to the shower axis.

The contour maps from different incident azimuth directions
in the case of pure geosynchrotron were studied in
\citet{2005APh....24..116H}. Besides elongation of field
strength pattern, they found total field strength pattern rotates
with the azimuth angle. The measurements of the individual
polarization components can be used to verify directly the geosynchrotron
origin of the signal in radio emission.

\clearpage

With the inclusion of the
charge excess and creation/destruction effect, the situations
become more complicated. As shown in Fig.~\ref{fig:E_contour_azimu},
the contours of the total field strength
show concentric ellipses, while there are some slight changes in the
orientation of these ellipses, it is not very obvious. This is not
surprising, for with both the geosynchrotron and creation/destruction emission
at work, the geometric relation is more complicated. Again, the NS and vertical
polarizations show more irregularity, in some cases with bivalve pattern.

\section{Conclusion}
\label{con}

In this paper, we describe our new Monte Carlo simulation of the
radio signal emitted by cosmic ray extensive air showers. Our basic approach
is similar to \emph{REAS2}, using Monte Carlo to generate a sample of particles
and calculate the field produced by them, but we included the charge excess effects
in addition to the geosynchrotron radiation. We use step functions in the
retard potentials to express the creation and destruction
of particles. At low frequency, the
radiation field can be derived classically.
The algorithm of our numerical program is presented
which has passed preliminary checks and gives results which are consistent
with the ones obtained by others.

We find that when the charge excess effect is included, the radio signal is significantly
modified: the magnitude of the signal is substantially reduced, and in the time domain
the pulse EW polarization exhibits a bipolar
pattern. This is the most important distinction to the previous pure
geosynchrotron radiation.
The charge excess effect on the frequency spectrum
is also considered. The geosynchrotron and
charge excess effect, when computed individually, have similar spectra which
drops at $\sim 100$MHz due to the loss of coherence. At low frequencies both have
flat spectrum, but when added the two tends to cancel each other and the spectrum
drops also at the lower end. These findings are in good agreement with recent results
reported in the literature \citep{2012NIMPA.662S.179H}.
We also computed the spectra at different locations. Near the
center, the charge excess effect amends the steep spectrum predicted by the
pure geosynchrotron mechanism, and the combined spectrum is in agreement with
the observation. Off the center, the theoretical spectrum is steeper than the
observation. This may be due to experimental error, or another kind of radiation, such as
the Cerenkov radiation.

We further apply our program to study the features of the signals. For a vertical
shower and near the shower axis, we find that the signal at any time comes from a wide
span of different heights, and indeed the layers from different heights
could all give sizeable contributions to the
total signal, though the shower maximum contributed most.
Far off center, the contribution from the shower
maximum dominates, while near the center,
the lower altitude layers could also contribute a large share.

We study the elevation dependence of the signal. At the shower center, as the
altitude raises, the peak magnitude increases slightly, then drops off if the altitude
reaches beyond that of the shower maximum. Off center,
it always decreases with increasing altitude.
This means that there is slight advantage to place the radio array detector
at sites of high altitude, if the array is designed to detect the radio signal
at center. On the other hand, if the array is sufficiently sensitive to be able
to detect the radio emission at large off-center distance, then there is not much
advantage to place it at high altitudes. Indeed, even in the former case, the
advantage is far less obvious than the particle-based detector.
We use an exponential function to fit the radial distributions and
find $R_0$ is about $100$ meters, which is consistent with experimental results.

We then consider the inclined showers with different zenith and azimuth angles, and
computed spatial distribution of the signal.
We find that the contour lines of signal strength are
basically concentric ellipses,
but due to the asymmetry of charges, there is an azimuthal asymmetry
in the EW polarization even for a vertically-downward air shower.
The total field strength and the dominant EW polarization
decrease gradually as the zenith angle increases, and the ellipses are elongated.
However, the addition of the charge excess effect obscured the signature
of the geosynchrotron effect, the
azimuthal dependence of the total field strength is not apparent.
In the NS polarization, the
shape is also changed from pintongs to bivalve pattern. Our program
could be a useful tool for incoming radio detection study.

This paper presents a very basic model of cosmic ray air shower radio emission, and
some similar results have been obtained previously. Nevertheless, it is useful to
verify these results with an independent computation as we did, and to examine how the signal
varies with elevation, shower inclination and azimuth angle, etc. Moreover,
it is a first step toward an independent, comprehensive numerical study of the
air shower radio emission problem.
We are working to improve our model by incorporating more physics effects and
implement more realistic models. We plan to use a shower model
generated by a modern Monte Carlo code (e.g. CORSIKA).
The effect from the variation of atmospheric refractive index
and the corresponding \v{C}erenkov radiation will also be investigated in
our subsequent works. We can then investigate showers of different energy and
composition, and then it will be useful to the radio detection experiments.

\section*{Acknowledgement}
\label{ack}
We would like to thank Tim Huege for providing us the Reas2.59
 code for comparison. This work
is supported by the Ministry of Science and Technology 863 project 2012AA121701,
by the Chinese Academy of Science Strategic Priority Research Program ``The Emergence of
Cosmological Structures''
of the Chinese Academy of Sciences, Grant No. XDB09000000,  and
the NSFC grant 11373030.

\appendix
\section{Coordinate Transformation}
\begin{figure}
\begin{center}
\includegraphics[width=.6\textwidth]{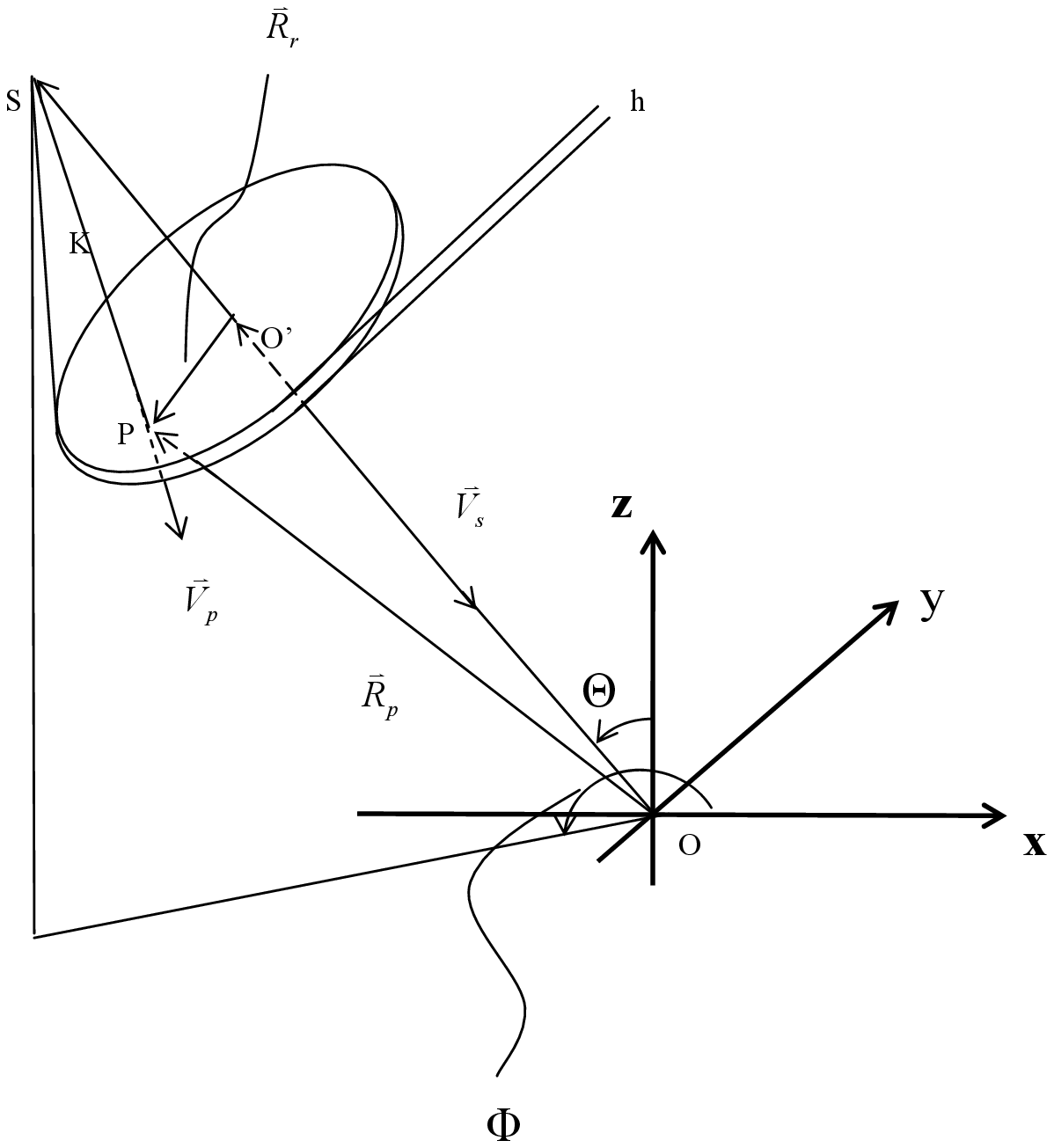}\\
\includegraphics[width=.6\textwidth]
{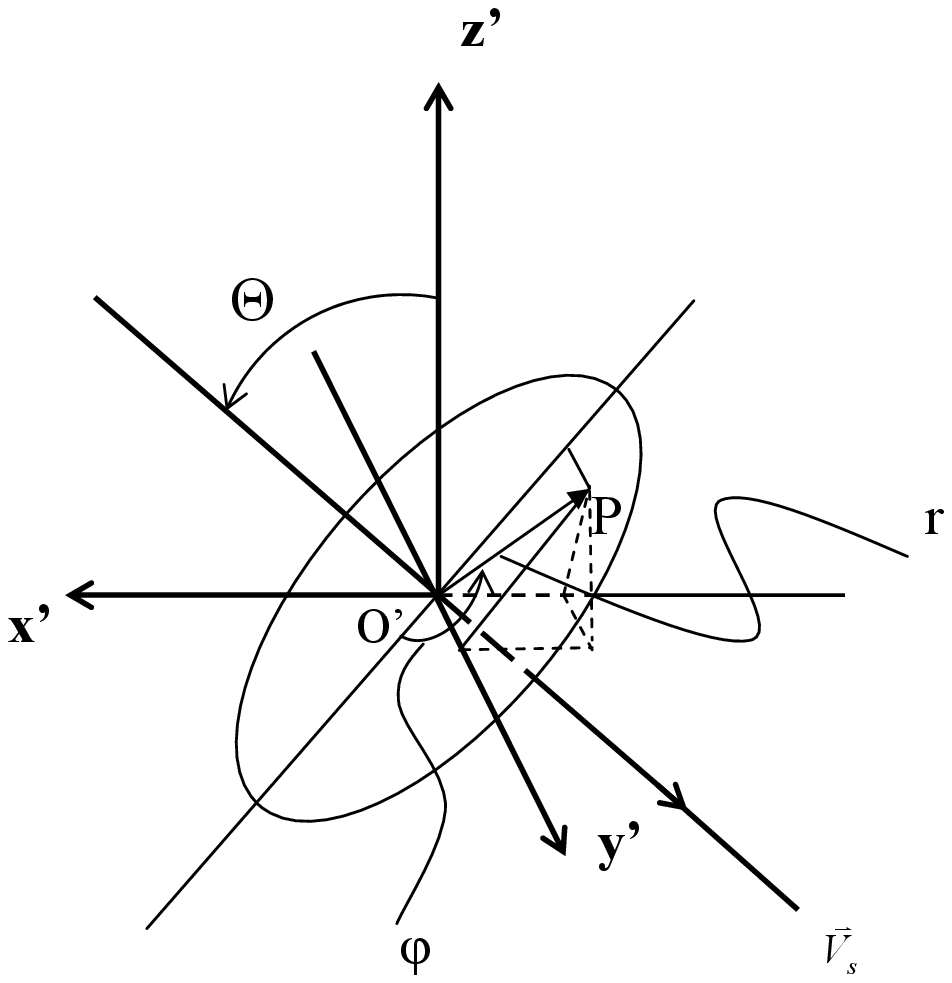}
\caption{
\label{fig:coordinate}
Top: a sketch of shower disk in Earth's coordinate system $XYZ$, with $X$ and $Y$
respectively pointing to the east and north. Down:  shower disk in local system $X'O'Y'$.
}
\end{center}
\end{figure}

In this part, the position and velocity of a charged particle in the Earth reference frame
are related to its relative position in the shower disk. In an incident shower
with zenith angle $\Theta$, the center of the disk plane is set to be the origin $O'$ of
the (moving) shower coordinates. The $X'$ axis is in the plane of the incidence,
pointing horizontally outwards, and $Y'$ axis is in the disk plane and normal to
the $X'$ axis with right hand side direction. The $Z'$ points upward (see the
right panel of Fig. \ref{fig:coordinate}). The relative position of the
particle $P$ can be written as
\begin{equation}
\vec{R}'_r =
 \begin{cases}
	r \cos \varphi \cos \Theta\; , \\
	r \sin \varphi\; , \\
   -r \cos \varphi \sin \Theta\; ,
 \end{cases}
\end{equation}
where $r$ is the distance from the origin $O'$ and $\varphi$ is the
azimuthal angle around the disk plane which rotates counterclockwisely
from lower part of shower disk.

The coordinate $\vec{R}_r$ of $P$ can be transformed from the $X'O'Y'$ to the
system $XO'Y$,
\begin{align}
x =& x' \cos \Phi - y' \sin \Phi\; , \\
y =& y' \cos \Phi + x' \sin \Phi\; .
\end{align}
Here in the $XO'Y$ plane, the $X$ and $Y$ axis respectively point to the
east and north(see left figure of \ref{fig:coordinate}).
$\Phi$ is the azimuthal angle in the system $XOY$.

The position of $O'$ in the $XYZ$ system, where the impact center $O$ is set to be
the origin, is given by
\begin{equation}
\vec{R}_{O'} = \vec{R}_{\mathrm{sf}} \times(R_{\mathrm{sf}} +d)/R_{\mathrm{sf}}\; ,
\end{equation}
where $\vec{R}_{\mathrm{sf}}$ is the position of center of shower front in the $XYZ$ system,
which is equal to $\vec{R}'_{\mathrm{sf}} -H\vec{\hat{e}}_{z}$, if the
impact center is $H$ meters above the sea level. Position $\vec{R}'_{\mathrm{sf}}$
could be evaluated according to the relation between height and atmospheric depth $X$
which is produced by a random number generator in the Monte Carlo code(see
subsection \ref{distr}), $d$ is the distance from the shower front. The
position in system $XYZ$ can be further obtained from the vector relationship
\begin{equation}
\vec{R}_{\mathrm{p}} = \vec{R}_{O'} + \vec{R}_{r}\; .
\end{equation}
Finally the absolute position is
\begin{equation}
\vec{R}_{\mathrm{p}} += H\ \vec{\hat{e}}_{z}\; .
\end{equation}

Secondary particles are assumed to be distributed within the spherical shell,
with radius $K$ equal to $2300$ m. Therefore their initial velocity direction
is assumed to be radial, i.e.
\begin{equation}
\vec{\hat{V}}_{\mathrm{p}} = \frac{\vec{R}_{\mathrm{p}} -\vec{R}_s}{|\vec{R}_{\mathrm{p}} -\vec{R}_s|}\; ,
\end{equation}
where
\begin{equation}
\vec{R}_s = \vec{R}_{O'} \times(R_{O'} + K)/R_{O'}\; .
\end{equation}

\section{The Motion of a charged particle in magnetic field}
The motion of a single charged particle in a static uniform magnetic field
is determined by the Lorentz equation
\begin{equation} \label{cross_equ}
\gamma m \frac{\dif\vec{V}}{\dif t} = -q \vec{V}\times \vec{B}\; ,
\end{equation}
where $\gamma$ is the Lorentz factor. Cross product $\vec{B}$ on both sides,
differentiating it, use Eq.(\ref{cross_equ}) and the vector identify
$\vec{B}\times (\vec{V}\times \vec{B}) = \vec{V}B^2 - \vec{B}(\vec{V}\cdot \vec{B})$\; ,
we get a second-order differential vector equation
\begin{equation} \label{vec_equ}
\frac{\dif^2 \vec{V}}{\dif t^2} + \left( \frac{q\vec{B}}{\gamma m}\right)^2 \vec{V}
- \left( \frac{q}{\gamma m}\right)^2 \vec{B}(\vec{V}\cdot \vec{B}) = 0\; .
\end{equation}
By dot-producting $B$ in both sides of the Lorentz equation,
\begin{equation}
\frac{\dif(\vec{V}\cdot \vec{B})}{dt} = 0\; ,
\end{equation}
i.e. $\vec{V}\cdot \vec{B} = \mathrm{const}$, so the solution of
Eq.(\ref{vec_equ}) is
\begin{equation} \label{V_equ}
\vec{V}(t) = \vec{a}_1\cos \omega_B t + \vec{b}_1\sin \omega_B t
+ \frac{\vec{B}(\vec{V}_0\cdot \vec{B})}{B^2}\; ,
\end{equation}
where $\omega^2_B = \left( q\vec{B}/\gamma m \right)^2$ is the gyration frequency
of the circular motion, and $\vec{V}_0$ is the initial velocity. The constants
$\vec{a}_1$ and $\vec{b}_1$ can be determined from the initial conditions,
\begin{equation}
\vec{a}_1 = \vec{V}_0 - \frac{\vec{B}(\vec{V}_0\cdot \vec{B})}{B^2}\; , ~~~~~~
\vec{b}_1 = \frac{q(\vec{a}_1 \times \vec{B})}{\gamma m\omega_B}\; .
\end{equation}
The trajectory of the particle can then be integrated, which is given by
\begin{equation}
\vec{X}(t) = \frac{\vec{a}_1}{\omega_B} \sin \omega_B t - \frac{\vec{b}_1}{\omega_B}(\cos \omega_B t - 1) + \frac{\vec{B}(\vec{V}_0\cdot \vec{B})}{B^2} t + \vec{X}_0\; ,
\end{equation}
where $\vec{X}_0$ is the initial position.

\bibliographystyle{raa}
\bibliography{REFRESH_reference}
\end{document}